\documentclass[twocolumn,preprintnumbers,amsmath,amssymb]{revtex4}
\usepackage{graphicx}% Include figure files
\usepackage{dcolumn}% Align table columns on decimal point
\usepackage{bm}% bold math
\usepackage{amsmath}
\usepackage{hyperref}

\def\epee{\mathrel{{\epsilon_{ee}}}}
\def\epem{\mathrel{{\epsilon_{e\mu}}}}
\def\epmm{\mathrel{{\epsilon_{\mu\mu}}}}
\def\epet{\mathrel{{\epsilon_{e\tau}}}}
\def\epmt{\mathrel{{\epsilon_{\mu\tau}}}}
\def\eptt{\mathrel{{\epsilon_{\tau\tau}}}}

\def \gta {\mathrel{\vcenter{\hbox{$>$}\nointerlineskip\hbox{$\sim$}}}}

\begin{document}

\preprint{LA-UR-04-1159}

\title{Solar neutrinos as probes of neutrino-matter interactions}% Force line breaks with \\

\author{Alexander Friedland}
 \email{friedland@lanl.gov}
\affiliation{%
Theoretical Division, T-8, MS B285, Los Alamos National
Laboratory, Los Alamos, NM 87545
}%

\author{Cecilia Lunardini}
 \email{lunardi@ias.edu}
\author{Carlos Pe\~na-Garay}
 \email{penya@ias.edu}
\affiliation{ School of Natural Sciences, Institute for Advanced
Study, Einstein Drive, Princeton, NJ 08540
}%

\date{February 25, 2004}% It is always \today, today,

\begin{abstract}
Data from solar neutrino and KamLAND experiments have led to a
discovery of nonzero neutrino masses. Here we investigate what
these data can tell us about neutrino interactions
with matter, including the poorly constrained flavor-changing
$\nu_e-\nu_\tau$ interactions. We give examples of the interaction
parameters that are excluded by the solar/KamLAND data and are beyond
the reach of other experiments.
%We describe the physical effects of
%such interactions on solar neutrino evolution and discuss
%the bounds on neutrino-matter interactions that can be inferred
%from current data.
We also demonstrate that
flavor-changing interactions, at the allowed level, may
profoundly modify the conversion probability for neutrinos of
energy $\lesssim 6$ MeV and the values of the mass parameter
inferred from the data. The implications for future experiments
are discussed.
\end{abstract}

\pacs{Valid PACS appear here}% PACS, the Physics and Astronomy
\maketitle

\section{Introduction}
\label{sec:intro}

For several decades, experiments have been trying to test the
Standard Model (SM) paradigm that neutrinos are massless and
interact only via the $W$ and $Z$ gauge boson exchange.
%The discovery of nonzero neutrino masses by the solar neutrino and
%KamLAND experiments has been a great triumph in particle physics.
In recent years, a breakthrough has emerged: data from the solar,
atmospheric, and reactor neutrino experiments have indicated that
neutrinos do have masses, and hence the SM is incomplete.
%At the moment, this
%remarkable result stands as the only solid evidence for physics
%beyond the SM.
It is the right time to ask whether the SM predictions for the
neutrino-matter interactions can be similarly tested.

The aim of this Letter is to investigate what can be learned about
neutrino-matter interactions from present and future solar and
KamLAND neutrino data. We answer two questions: (i) Can the solar
and KamLAND experiments constrain parts of the parameter space
that are presently inaccessible by non-oscillation experiments?
(ii) Can the uncertainty in our present knowledge of
neutrino-matter interactions affect the determination of the
oscillation parameters? As we show, the answer to both questions
is affirmative. We give explicit examples of parameters that are
disfavored by solar and KamLAND data and that are beyond the reach
of non-oscillation experiments. We also demonstrate that
nonstandard interactions (NSI), at an allowed level, can
qualitatively modify the fit to the data and change the values of
inferred mass parameters. This scenario leads to nontrivial
predictions for future experiments.
%We focus on the physical
%effect of the NSI on the solar neutrino survival probability and
%the main conceptual results.
A full presentation of the numerical constraints we obtain is
beyond the scope of this Letter and will be given elsewhere
\cite{inprep}.

We study the effects of NSI \emph{on top of} nonzero neutrino
masses and mixing, since the latter are required by the KamLAND
data. This scenario has been previously studied
\cite{Roulet,Barger:1991ae,Fogli:1994xv,Bergmann:1998mr} with an
emphasis on the limit of small flavor-changing interactions. We
extend the formalism developed in these papers to make it
applicable to our problem.

\section{NSI and solar neutrinos: the physics}

Low-energy
%($E_\nu \ll M_W, M_Z$)
neutrino interactions can be described by four-fermion interaction
vertices, $L \ni \sum \bar\psi\psi\bar\psi\psi$.
%A vertex could have one, two, or four
%neutrino lines. The four-neutrino vertices are not relevant to
%solar or reactor neutrinos and are not discussed in this paper.
%The vertices with a single neutrino line are relevant for the
%neutrino production in the Sun and reactors. They are well
%constrained experimentally, and their measured properties agree
%with the predictions of the SM for charged current (CC)
%interactions. Finally, the vertices involving two neutrinos
%($2\nu$) are relevant for neutrino propagation in matter. While
%the non-oscillation experiments provide good constraints for some
%of them, others are rather poorly known. These vertices will be
%our main focus.
The vertices affecting neutrino evolution in matter are those
containing two neutrino lines ($2\nu$). In the SM, these vertices
receive contributions from neutral current (NC) processes and, if
the initial state contains a charged lepton, also charged current
(CC) processes. The NC processes are predicted to be
flavor-preserving and universal.
%How well have these assertions been tested?
Possible non-standard (both flavor-preserving and flavor-changing)
contributions to the $2\nu$ vertices can, most generally, be
parameterized as
\begin{eqnarray}
L^{NSI} &=& - 2\sqrt{2}G_F (\bar{\nu}_\alpha\gamma_\rho\nu_\beta)
(\epsilon_{\alpha\beta}^{f\tilde{f} L}\bar{f}_L \gamma^\rho
\tilde{f}_L + \epsilon_{\alpha\beta}^{f\tilde{f}
R}\bar{f}_R\gamma^\rho \tilde{f}_{R})\nonumber\\
&+& h.c. \label{eq:lagNSI}
\end{eqnarray}
Here $\epsilon_{\alpha\beta}^{f\tilde{f} L}$
($\epsilon_{\alpha\beta}^{f\tilde{f} R}$) denotes the strength of
the NSI between the neutrinos $\nu$ of flavors $\alpha$ and
$\beta$ and the left-handed (right-handed) components of the
fermions $f$ and $\tilde{f}$; $G_F$ is the Fermi constant.

%Remarkably, direct experimental bounds on some of the epsilons are
%rather poor.
Bounds on the epsilons come from accelerator-based
experiments, such as NuTeV \cite{nutev}  and CHARM \cite{charm},
and experiments involving charged leptons. In the later case, we
do not include bounds obtained by the $SU(2)$ symmetry, since
strictly speaking these can be avoided if, for example, the
corresponding operators contain Higgs doublets
\cite{BerezhianiRossi}. Both types of experiments are quite
effective at constraining the vertices involving the muon
neutrino, giving $\epem\lesssim 10^{-3}$, $\epmm\lesssim
10^{-3}-10^{-2}$. At the same time, bounds on $\epee$,
$\epet$, and $\eptt$ are rather loose, {\it e.g.},
$|\epsilon_{\tau\tau}^{uu R}|<3$, $-0.4<\epsilon_{ee}^{uu R}<0.7$,
$|\epsilon_{\tau e}^{uu}|<0.5$, $|\epsilon_{\tau e}^{dd}|<0.5
$\cite{Davidson:2003ha}.

NSI can modify both the neutrino propagation (oscillation)
\cite{W} and neutrino detection processes. The propagation effects
of NSI are, first of all, only sensitive to
$\epsilon_{\alpha\beta}^{f\tilde{f}}$ when $f=\tilde{f}$
\footnote{Among other reasons, the processes changing the flavor
of the background fermion do not add up coherently \cite{nunu1}.}
(henceforth,
$\epsilon_{\alpha\beta}^{ffP}\equiv\epsilon_{\alpha\beta}^{fP}$),
and, second, only to the vector component of that interaction,
$\epsilon_{\alpha\beta}^{f}\equiv\epsilon_{\alpha\beta}^{fL}+\epsilon_{\alpha\beta}^{fR}$.
The matter piece of the oscillation Hamiltonian can be written (up
to an irrelevant constant) as:
\begin{equation}
H_{\rm mat}^{3\times3}=\sqrt{2}G_F n_e\begin{pmatrix}
1+\epee & \epsilon_{e\mu}^\ast & \epsilon_{e\tau}^\ast \\
\epem & \epmm& \epsilon_{\mu\tau}^\ast \\
\epet & \epmt & \eptt \\
\end{pmatrix},
\label{eq:ham}
\end{equation}
where $n_e$ is the number density of electrons in the medium. The
epsilons here are the sum of the contributions from electrons
($\epsilon^{e}$), up quarks ($\epsilon^{u}$), and down quarks
($\epsilon^{d}$) in matter: $\epsilon_{\alpha\beta}\equiv
\sum_{f=u,d,e}\epsilon_{\alpha\beta}^{f}n_f/n_e$. Hence,
unlike in the standard case
($\epsilon_{\alpha\beta}=0$), the NSI matter effects depend on the
chemical composition of the medium.
%, and not only on the electron density $n_e$.

The CC detection reactions at SNO, KamLAND and the radiochemical
experiments, just like the production reactions in the Sun, are
unchanged by Eq.~(\ref{eq:lagNSI}). On the other hand, the
neutrino-electron elastic scattering (ES) reactions at
Super-Kamiokande and SNO, and the NC reaction at SNO could be
affected.
%The cross sections of these
%reaction depend not only on the vector, but also on the axial part
%of the interaction.
The SNO NC reaction is an axial current process
\cite{Donnelly:1974pq}, while the ES reaction depends on both axial
and vector parts. Hence, the former is independent of the
oscillation Hamiltonian (\ref{eq:ham}), while the latter is not.

Since both $\epem$ and $\epmm$ are strongly constrained, we
set them to zero and vary $\epee$, $\epet$, $\eptt$. Even with this
reduction, the parameter space of the problem is quite large:
different assignment of the diagonal and offdiagonal NSI to electrons
and $u$ and $d$ quarks yield different dependences of the oscillation
Hamiltonian on the chemical composition and different detection cross
sections. To avoid complicating our main point with technical
details, we limit our study to the case of NSI on quarks, assigning
the same strength to the neutrino interactions with $u$ and $d$ quarks.

For the solar neutrino analysis, we perform the standard reduction
of the $3\times 3$ Hamiltonian to a $2\times 2$ Hamiltonian
\cite{3nu}. This involves performing a rotation in the $\mu-\tau$
subspace by the atmospheric angle $\theta_{23}$ and taking  the
first two columns/rows. This simplification is valid if (i) the
$1-3$ mixing angle is small: $\theta_{13}\ll 1$ and (ii)
$G_Fn_e\epet\ll\Delta m^2_{\rm atm}/(2 E_\nu)$, with $E_\nu$ being
the neutrino energy and $\Delta m^2_{\rm atm}$ the difference of
the squared masses, $\Delta m^2_{\rm atm}\equiv m^2_3-m^2_2$, as
given by atmospheric neutrino data.  The first requirement is
ensured by the experimental bound from CHOOZ
\cite{Apollonio:1998xe}; the second one can be checked to hold
even for $\epet$ of order unity. The vacuum oscillation
Hamiltonian then takes the usual form
\begin{eqnarray}
H_{\rm vac} = \left(\begin{array}{rr}
  -\Delta \cos 2\theta & \Delta \sin 2\theta \\
  \Delta \sin 2\theta & \Delta \cos 2\theta
\end{array} \right),
\label{eq:VAC_convention}
\end{eqnarray}
where $\Delta\equiv \Delta m^2/(4 E_\nu)$ and $\Delta m^2$ is the
mass splitting between the first and second neutrino mass states:
$\Delta m^2\equiv m^2_2-m^2_1$.
The matter contribution can be written (once again, up to an
irrelevant overall constant) as:
\begin{eqnarray}
H_{\rm mat}^{NSI} = \frac{G_F n_e}{\sqrt{2}}
\left(\begin{array}{cc}
   1+\epsilon_{11} & \epsilon^\ast_{12} \\
  \epsilon_{12}   & -1-\epsilon_{11}
\end{array} \right),
\label{eq:ceciconv}
\end{eqnarray}
where the quantities $\epsilon_{ij}$ ($i=1,2$) depend on the
original epsilons and on the rotation angle $\theta_{23}$:
\begin{eqnarray}
\epsilon_{11}=\epee - \eptt \sin^2 \theta_{23}~,~~~~~~~~
\epsilon_{12}=-2\epet \sin \theta_{23}~. \label{eq:cecidef2}
\end{eqnarray}
In Eq. (\ref{eq:cecidef2}), small corrections of order $\sin
\theta_{13}$ or higher have been neglected.
We introduce a useful parameterization:
\begin{eqnarray}
H_{\rm mat}^{NSI} = \left(\begin{array}{cc}
  A \cos 2\alpha & A e^{-2i\phi} \sin 2\alpha \\
  A e^{2i\phi} \sin 2\alpha  & -A \cos 2\alpha
\end{array} \right).
\label{eq:MAT_convention}
\end{eqnarray}
Here the parameters $A$, $\alpha$ and $\phi$ are defined as
follows:
\begin{eqnarray}
&&\tan 2\alpha = |\epsilon_{12}|/(1+\epsilon_{11}) ~,~~~~~
2\phi=Arg(\epsilon_{12}),\nonumber \\
%&&\cos 2\phi = \frac{\Re e({\epsilon_{12}})}{|\epsilon_{12}|}~,~~~~~
%\sin 2\phi =-\frac{\Im m({\epsilon_{12}})}{|\epsilon_{12}|}\nonumber \\
&&A= G_F n_e \sqrt{[(1+\epsilon_{11})^2+|\epsilon_{12}|^2]/2}~.
\label{eq:cecidef}
\end{eqnarray}
In absence of NSI we have $A=G_F n_e/\sqrt{2}$, $\alpha=0$ and the
Hamiltonian (\ref{eq:MAT_convention}) reduces to its standard
form.

Notice the appearance of the phase $\phi$ in
Eq.~(\ref{eq:MAT_convention}). Since the phases of the basis
states are chosen to make the elements in
Eq.~(\ref{eq:VAC_convention}) real, $\phi$ cannot be
\emph{simultaneously} removed. This has been noted in the studies
of terrestrial neutrino beams
\cite{Gonzalez-Garcia:2001mp,Campanelli:2002cc}, but overlooked in
the solar neutrino literature.

What is the physical range for the parameters $\theta$, $\alpha$
and $\phi$? In the standard case $\alpha=0$, the physical range of
$\theta$ is $[0,\pi/2]$, including the so-called ``light" and
``dark" sides \cite{darkside}. A generalization to the NSI case is
%\alex{think}
\begin{equation}\label{eq:range}
    \theta \in [0,\pi/2],\;\;\;\alpha \in [0,\pi/2],\;\;\;\phi \in
    [-\pi/2,\pi/2].
\end{equation}
These ranges of parameters cover all possibilities in
Eqs.~(\ref{eq:VAC_convention},\ref{eq:MAT_convention}).
%We will show that,
For solar neutrinos, the range of $\phi$ could be cut in half,
since points with $\phi$ and $-\phi$ give the same probability
$P_{ee}$ that a $\nu_e$ produced in the Sun is seen as a $\nu_e$
in a detector. Moreover, the points $(\theta, \alpha, \phi)$ and
$(\pi/2-\theta, \pi/2-\alpha, \phi)$ are related by
$P_{ee}\leftrightarrow P_{\mu\mu}$, which are equal in the
2-neutrino case by unitarity.

%It can be checked that, within this range, the points $(\theta,
%\alpha, \phi)$ and $(\pi/2-\theta, \pi/2-\alpha, \phi)$
%%\alex{check}
%give the same survival probability for solar neutrinos.

Let us determine the expression for $P_{ee}$. We first note that,
because KamLAND selects $\Delta m^2\gtrsim 10^{-5}$ eV$^2$,
coherence between the Hamiltonian eigenstates is completely lost
once one integrates over the neutrino energy spectrum and the
neutrino production region in the Sun. The expression for the
incoherent survival probability can be most easily
derived in the basis that diagonalizes the matter Hamiltonian
(\ref{eq:MAT_convention}). We obtain the familiar form \cite{Parke}
\begin{eqnarray}\label{eq:Pee_master}
  P_{ee} = [1+(1-2P_c)\cos 2\theta_\odot\cos 2\theta]/2,
\end{eqnarray}
where $\theta_\odot$ is the mixing angle at the neutrino
production point in the solar core and $P_c$ is the level crossing
probability. These quantities contain all the effects of the NSI.
The angle $\theta_\odot$ is given by
\begin{eqnarray}\label{eq:cos2theta1}
    \cos 2\theta_\odot &=& \frac{\cos 2\theta -x_\odot\cos 2\alpha}
{\sqrt{1+x_\odot^2+2 x_\odot\cos 2\theta_{\rm rel}}},\\
\cos 2\theta_{\rm rel} &\equiv& \sin 2\theta\sin 2\alpha \cos
2\phi - \cos 2\theta \cos 2\alpha.
\end{eqnarray}
Here $x_\odot\equiv A/\Delta$ at the neutrino production point.

The expression for $P_c$ is also easily found in the same basis,
where it becomes apparent that the dynamics of conversion in
matter depends only on the relative orientation of the eigenstates
of the vacuum and matter Hamiltonians. This allows to directly apply
the known analytical solutions for $P_c$, and, upon
rotating back, obtain a generalization of these results to the NSI
case. For example, the answer for the infinite exponential profile
\cite{Toshev1987,Petcov1988} $A\propto \exp(-r/r_0)$ becomes
$P_c=(\exp[\gamma(1-\cos 2\theta_{\rm
rel})/2]-1)/[\exp(\gamma)-1]$, where $\gamma\equiv 4\pi r_0
\Delta=\pi r_0 \Delta m^2/E_\nu$.
%This expression by itself,
%however, is not very useful for our case, since for $\Delta
%m^2\gtrsim 10^{-5}$ eV$^2$ the solar profile cannot be thought as
%an infinite exponential. On the other hand,
We further observe that since $\gamma\gg 1$ the adiabaticity
violation occurs only when $|\theta-\alpha|\ll 1$ and $\phi
\simeq\pi/2$, which is the analogue of the small-angle MSW
\cite{W,MS} effect in the rotated basis. The ``resonant''
region in the Sun where level jumping can take place is narrow,
defined by $A\simeq\Delta$ \cite{jumpinginsideout}. A neutrino
produced at a lower density evolves adiabatically, while
a neutrino produced at a higher density may undergo level crossing.
The probability $P_c$ in the latter case is given to a very good accuracy by the formula
for the linear profile,
%(which is a limit of the exponential \cite{jumpinginsideout}),
with an appropriate gradient taken along the neutrino trajectory,
%\begin{equation}\label{eq:Pc_full}
%    P_c\simeq\Theta(A-\Delta)e^{\gamma (\cos 2\theta \cos 2\alpha-\sin 2\theta\sin 2\alpha
%    \cos 2\phi-1)/2},
%\end{equation}
\begin{equation}\label{eq:Pc_full}
    P_c\simeq\Theta(A-\Delta)e^{-\gamma (\cos 2\theta_{\rm rel}+1)/2},
\end{equation}
where $\Theta(x)$ is the step function, $\Theta(x)=1$ for $x>0$
and $\Theta(x)=0$ otherwise. We emphasize that our results differ
from the similar ones given in
\cite{Mansour:1998fi,Bergmann:1998mr} in three important respects:
(i) they are valid for all, not just small values of $\alpha$
(which is essential for our application), (ii) they include the
angle $\phi$, and (iii) the argument of the $\Theta$ function does
not contain $\cos 2\theta$, as follows from
\cite{jumpinginsideout}. We stress that for large values of
$\alpha$ and $\phi\simeq\pi/2$ adiabaticity is violated for large
values of $\theta$.

Finally, to get an idea on the size of the day/night asymmetry,
$A_{\rm DN}\equiv 2 (N-D)/(N+D)$, (here $D$ ($N$) denotes the
$\nu_e$ flux at the detector during the day (night)) we can model
the Earth as a sufficiently long (compared to the oscillation
length) object of constant density. For $^8$B neutrino energies,
this is appropriate for $\Delta m^2 \gta 3-5 \cdot 10^{-5}~{\rm
eV^2}$. Introducing a small parameter $x_\oplus\equiv A/\Delta$,
where $A$ is evaluated for a typical density inside the Earth, we
find, to the first order in $x_\oplus$,
%  In this model, we find
\begin{equation}\label{eq:A_dn_phi}
    A_{\rm DN} \simeq x_\oplus\frac{\sin 2\theta(\cos
      2\alpha \sin 2\theta + \cos 2\phi \sin 2\alpha \cos 2\theta)}
{-[\cos 2\theta_\odot (1-2P_c)]^{-1}
      - \cos 2\theta}.
\end{equation}
We verified that Eq.~(\ref{eq:A_dn_phi}) gives a good agreement with
precise numerical calculations for $n_e \simeq 1.6$ moles/cm$^3$.
For the lower $\Delta m^2$ region allowed by KamLAND, $\Delta m^2
\gta 1 -3 \cdot 10^{-5}~{\rm eV^2}$, the oscillation length is
comparable to the size of the Earth, however, the averaging in Eq.
(\ref{eq:A_dn_phi}) still applies to a signal integrated over the
zenith angle.

\begin{figure}[htbp]
  \centering
  \includegraphics[width=0.47\textwidth]{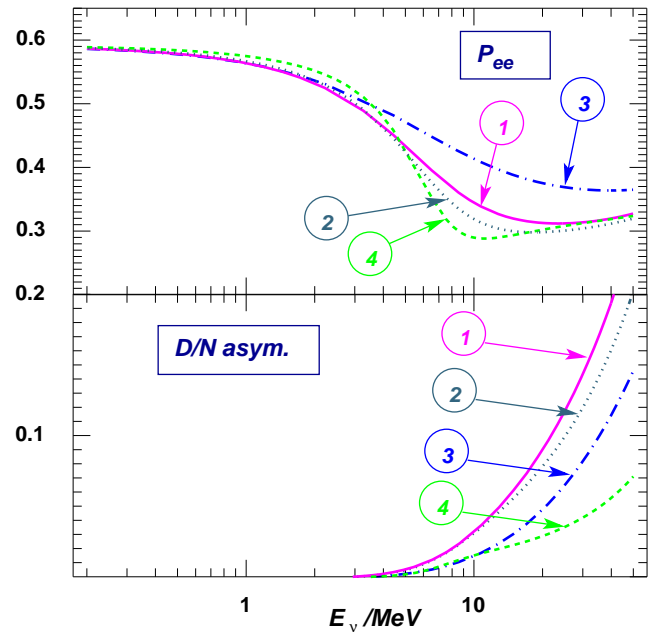}
  \caption{The electron neutrino survival probability and the day/night
  asymmetry as a function of energy for  $\Delta m^2=7\times 10^{-5}$ eV$^2$,
  $\tan^2\theta=0.4$ and several representative values of the NSI parameters: (1)
  $\epsilon_{11}^{u}=\epsilon_{11}^{d}=\epsilon_{12}^{u}=\epsilon_{12}^{d}=0$;
  (2) $\epsilon_{11}^{u}=\epsilon_{11}^{d}=-0.008$,
  $\epsilon_{12}^{u}=\epsilon_{12}^{d}=-0.06$;
  (3) $\epsilon_{11}^{u}=\epsilon_{11}^{d}=-0.044$,
  $\epsilon_{12}^{u}=\epsilon_{12}^{d}=0.14$;
  (4) $\epsilon_{11}^{u}=\epsilon_{11}^{d}=-0.044$,
  $\epsilon_{12}^{u}=\epsilon_{12}^{d}=-0.14$.
  Recall that the parameters in Eq. (\ref{eq:cecidef2}) equal
  $\epsilon_{ij}=\epsilon^u_{ij}n_u/n_e+\epsilon^d_{ij}n_d/n_e$.}
  \label{fig:PeeAdn}
\end{figure}

In Fig. \ref{fig:PeeAdn} we plot the neutrino survival probability
as a function of energy for several representative values of the
NSI parameters. We take $\Delta m^2$ and $\theta$ corresponding to
the best-fit LMA point and choose the production point to be at
$r=0.1 R_\odot$. Curve (1) is the standard interaction case, given
for reference. The other three curves represent the three
qualitatively different regimes that are of interest to us. In the
following we illustrate them in connection with observations. For
definiteness, we consider real values of $\epsilon_{12}$, both
positive ($\phi=0$) and negative ($\phi=\pi/2$). As is clear
from Eq. (\ref{eq:MAT_convention}), complex values
($0<\phi<\pi/2$) interpolate between these two cases.

\section{Analysis of data}

We now turn to the comparison of the NSI predictions with
observations. To do this, we perform a best fit analysis of the
solar neutrino and KamLAND data along the lines of
Refs.~\cite{solarfit,klandfit}.
%
%The fit in question is performed to the solar neutrino and KamLAND data.
In particular, solar data include the radiochemical rates
~\cite{chlorine,sage,Hampel:1998xg,gno}, the SK ES
zenith-spectra~\cite{sksollast}, the SNO day-night
spectra~\cite{Ahmad:2001an,Ahmad:2002jz,Ahmad:2002ka} measured in
phase-I and the SNO rates measured in phase-II~\cite{snoII}. For
consistency, the NC rate prediction for SNO is treated as a free
parameter because it is affected by an unknown change in the axial
coupling of the quarks that could accompany the vector NSI considered
in our analysis \cite{Vogel2002}. In our calculations, we
use the updated BP04~\cite{bp04} Standard Solar Model (SSM) fluxes,
electron density and neutrino production point distributions in the
Sun. For KamLAND we considered the measured antineutrino spectrum with
visible energies higher than 2.6 MeV \cite{kamland}.

The key ingredients of our analysis turn out to be the rates and
energy spectrum data from SNO and Super-Kamiokande. A comparison
of the SNO CC rate with the Super-Kamiokande rate
\cite{Ahmad:2001an} and the SSM indicates that, within the energy
range accessible for the two experiments, the electron neutrino
survival probability is about 30\%.  No other distinguishing
features, such as a day/night asymmetry or spectral distortion,
are seen at a statistically significant level \cite{Ahmad:2002ka}.
In the case of the SM interaction, this turn out to be a very
restrictive condition;
%there are only two relatively small regions in the
%2-neutrino parameter space, the so-called LMA and LOW solutions, that
%have this behavior. KamLAND \cite{kamland} rules out the LOW solution
%and further constrains the LMA solution.
as seen in Fig.~\ref{fig:PeeAdn}, the range of energies for which
the survival probability is constant at 30\% (henceforth, ``the
flat window'') is barely large enough to cover the SNO energy
window. On the low-energy end, the resonant condition in the solar
core increases the neutrino survival probability; on the
high-energy end, the resonant condition in the earth causes a
large D/N effect.  Hence, values of the NSI parameters that
``shrink" the flat window, or shift it in the region disfavored by
KamLAND, can be excluded. Conversely, if NSI increase the size of
the flat window, new solutions may emerge.

\subsection{$\epsilon_{12}>0$}

A typical behavior for this case is exhibited by curve (3): the
``step" in $P_{ee}$ becomes longer and the day/night asymmetry is
not much smaller than in the SM case. These features point to a
possible conflict with data.
%Indeed, a
%fit to data shows that the best-fit point has a significantly worse $\chi^2$
%than in the standard case.  \alex{quantify}
Our analysis confirms this expectation: a parameter scan
\cite{inprep} for $\epsilon_{12}>0$ reveals that a significant
fraction of the parameter space which is allowed by the
accelerator-based data can be excluded by the solar/KamLAND data.
As an example, we find that points with  $\epsilon_{11} =0$ and
$\epsilon^u_{12}>0.14 $ (here and later,
$\epsilon^u_{\alpha\beta}=\epsilon^d_{\alpha\beta}$ is assumed)
%\alex{move this to 3b;
%instead  give results for standard norm (our curve (3))}
%(corresponding to $\epet\simeq ...$, see Eq.
%(\ref{eq:cecidef2})) \alex{fill number in}
are unacceptable at $90\%$ confidence level (C.L.). If we keep $A$
in the core of the Sun (at $r=0.05 R_\odot$) fixed to its standard
value, $A=G_F n_e/\sqrt{2}$, we exclude points with
$\epsilon^u_{12}>0.11$ at the same C.L. (for 1 degree of freedom,
d.o.f, unless specified otherwise).  The accelerator experiments
allow values of order unity (in absolute value) for this parameter
\cite{Davidson:2003ha}. We stress that the latter probe only
$|\epsilon_{12}|$, while, as we show here, oscillation experiments
are sensitive to the complex phase $\phi$ (or, for real epsilons,
to the sign of $\epsilon_{12}$).
%\alex{think!}.

\subsection{$\epsilon_{12}<0$}

%We now turn to the case $\epsilon_{12}>0$. In this case, the
%effect on the survival probability is quite different, unlike the
%accelerator-based data, which is sensitive to $|\epsilon_{12}|$.

For  $\epsilon_{12}$ close to zero ($-0.08\lesssim\epsilon^u_{12}
< 0$),
%\alex{check}),
the only effect of the NSI is to flatten the part of the $P_{ee}$
curve around 5-6 MeV, as illustrated by curve (2) in Fig.
\ref{fig:PeeAdn}.  No new solutions appear and the allowed region in
the $\theta$-$\Delta m^2$ plane is similar to that
obtained with the SM interactions. This scenario has important
implications for SNO, which can probe it by lowering its energy
threshold.

\begin{figure}[htbp]
  \centering
  \includegraphics[width=0.47\textwidth]{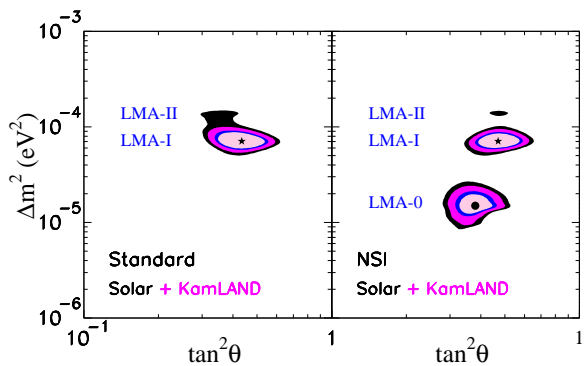}
  \caption{Regions of $\Delta m^2$ and $\tan^2\theta$ allowed at
$90,95,99,99.73\%$ C.L. (2 d.o.f) for SM interactions (left) and the NSI
scenario (right) described by Eqs.
(\ref{eq:VAC_convention}-\ref{eq:MAT_convention}). For the latter
we used $\epsilon_{11}^{u}=\epsilon_{11}^{d}=-0.065$,
$\epsilon_{12}^{u}=\epsilon_{12}^{d}=-0.15$.}
  \label{fig:fits}
\end{figure}

Finally, curve (4), obtained for $\epsilon^u_{11}=-0.044$,
$\epsilon^u_{12}=-0.14$, represents a novel and very interesting
physical possibility. Its main feature is a significantly wider flat
window, compared to the standard case. The key reason is the
suppression of the day/night asymmetry on the high-energy end of the
window. The physics of the suppression can be understood from
Eq.~(\ref{eq:A_dn_phi}), which, for $\phi=\pi/2$, gives $A_{DN}\propto
\sin(2\theta-2\alpha)$. If the parameters are chosen in such a way
that $\theta$ and $\alpha$ \emph{in the Earth} are comparable, the
Earth regeneration effect is suppressed. Because of the difference in
chemical composition, the difference $\theta-\alpha$ is larger in the
Sun and, consequently, the evolution in the Sun is still adiabatic.

A broader flat window allows the fit region to extend to lower
values of $\Delta m^2$
%(which shifts the curves in
%Fig.~\ref{fig:PeeAdn} to lower energies) by a factor of 3-4,
compared to the standard case. While KamLAND excludes the middle
part of the new region, the bottom part of that region ($\Delta
m^2 \sim (1-2)\times 10^{-5}$ eV$^2$) is, in fact, allowed
\cite{kamland,klandfit}. Thus, in addition to the usual two
solutions, LMA-I and LMA-II, a completely new disconnected
solution emerges, which we shall denote LMA-0.

\begin{figure}[htbp]
  \centering
  \includegraphics[width=0.47\textwidth]{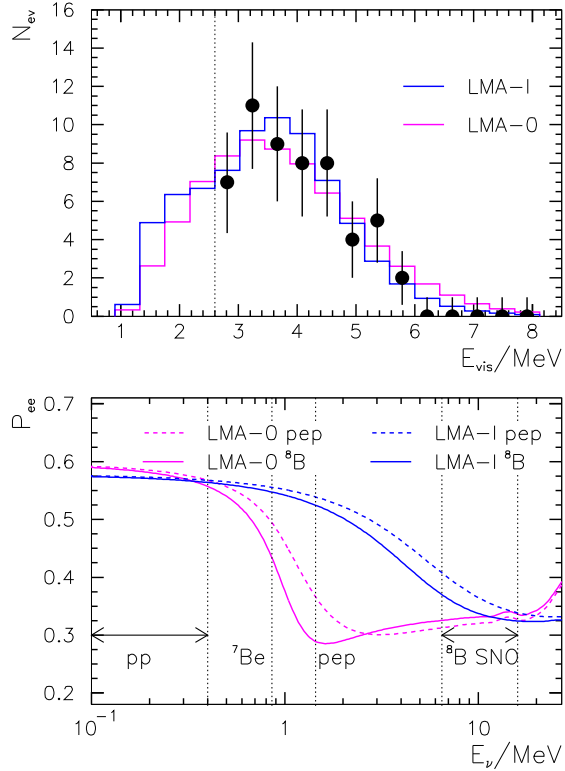}
  \caption{The predicted KamLAND spectrum (top) and the time-averaged
  solar neutrino survival probability (bottom) for the LMA-0 best-fit
  point. For comparison, the standard LMA-I survival probability is
  also given. Refer to the text for details.
%The curves for survival probabilities are averaged over
%  the production region for $^8$Be and pep neutrinos (see legend). The
%  dotted vertical lines indicate the intervals of energy relevant for
%  the different components of the solar neutrino flux. The interval for $^8$B neutrinos is cut from below to account for the SNO energy threshold.
}
  \label{fig:Pee}
\end{figure}

The situation is illustrated in Fig.~\ref{fig:fits}, in which we
compare the allowed regions in the standard case to those computed
for chosen values of the NSI parameters:
$\epsilon_{11}^{u}=-0.065$, $\epsilon_{12}^{u}=-0.15$.  The
best-fit point in the LMA-0 region has $\Delta m^2=1.5\times
10^{-5}$ eV$^2$ and $\tan^2\theta=0.39$, with $\chi^2=81.7$. For
the same NSI parameters, the $\chi^2$ has another minimum,
$\chi^2=79.9$, at $\Delta m^2=7.1\times 10^{-5}$ eV$^2$ and
$\tan^2\theta=0.47$, corresponding to the LMA-I solution.  The quality
of the fit for LMA-0 and LMA-I is comparable; if only the
KamLAND rate, and not spectrum, information is used, the LMA-0 fit
is slightly better ($\chi^2=73.0$, against $\chi^2=73.7$ for the
minimum in the LMA-I region). For comparison, the best fit
parameters for the standard case are
 $\Delta m^2=7.1\times 10^{-5}$ eV$^2$ and
$\tan^2\theta=0.43$, with $\chi^2=79.6$ (using the KamLAND
spectrum).
%\alex{replace empty words by numbers}

We stress that the existence of the LMA-0 solution depends mainly
on the value of $\alpha$ in the Earth and, to a lesser extent, on
the value of the norm $A$. Hence, this solution persists for other
choices of the NSI parameters, so long as they yield approximately
the same $\alpha$ in the Earth. For example, if the diagonal
interactions are assumed to be standard, $\epsilon_{11}^{u}=0$,
one finds a good fit in the LMA-0 region for
$\epsilon_{12}^{u}\simeq -0.25$.

Our LMA-0 solution should not, of course, be confused with the
``VERY-low-LMA" solution \cite{verylowlma}, which arises under
completely different physical assumptions, namely, if one assumes
large ($\sim 5-8$\%) density fluctuations \cite{Loreti1994} in the
Sun \footnote{The VERY-low-LMA solution has been recently shown to
be disfavored \cite{Burgess2003} by the salt-phase results from
SNO.}.

We note that the LMA-0 solution requires that the value of the
$\phi$ angle be not too different from $\pi/2$. Numerically, if we
fix all the other parameters to the values of fig. \ref{fig:fits}
and vary $\phi$, we find that LMA-0 disappears at 90\% C.L. for
$\phi<0.45\pi$.  As $\phi$ is decreased further, the goodness of
the overall fit decreases. Indeed, for $\phi=0$ (positive
$\epsilon_{12}$), the survival probability has the features of
curve (3) in Fig. \ref{fig:PeeAdn}, which are disfavored, as
discussed earlier. The LMA-I solution disappears at 90\% C.L. for
$\phi<0.31\pi$.  A scan over the region of $\epsilon_{12}$ real
and negative gives exclusion of regions of the parameter space
allowed by accelerator limits.  For instance, points with
$\epsilon_{11} =0$ and $\epsilon^u_{12}<-0.32$ are unacceptable at
$90\%$ C.L.  For $A$ fixed to the standard value $G_F
n_e/\sqrt{2}$ in the solar core, the limit is
$\epsilon^u_{12}<-0.19$, at $90\%$ C.L.

%We stress that the values of the epsilons required for our LMA-0
%solution are not at present excluded experimentally. Indeed,
Our
choice of $\epsilon^u_{12}=-0.15$ implies
$\epsilon^u_{e\tau}=\epsilon^{uL}_{e\tau}+
\epsilon^{uR}_{e\tau}\simeq 0.11$ (see Eq. (\ref{eq:cecidef2}), in
which we set $\theta_{23}=\pi/4$),
i.e., for example, $\epsilon^{uL}_{e\tau}\simeq
\epsilon^{uR}_{e\tau} \sim 0.05$.
%\alex{Check!}.
This is about one order of magnitude smaller than the direct bound
from CHARM \cite{Davidson:2003ha}. A more interesting question is
whether the NSI parameters of interest for the LMA-0 scenario could be
tested with atmospheric neutrinos. For our specific case, the
existing two-neutrino analyses \cite{Fornengo:2001pm,brazilians}
do not provide an answer, as the problem is essentially a
three-flavor one. Our investigation \cite{inprep} shows that
regions exist in the space of the NSI parameters where the effect
of NSI on the atmospheric neutrino observables is minimal and a
satisfactory fit to the data is obtained. As an example, a point
in this allowed region is
$\epsilon^u_{ee}=\epsilon^d_{ee}=-0.025$,
$\epsilon^u_{e\tau}=\epsilon^d_{e\tau}=0.11$,
$\epsilon^u_{\tau\tau}=\epsilon^d_{\tau\tau}=0.08$.
%\alex{to be checked again}.
%(\alex{give epsilons}).

The survival probabilities for the best-fit point of the LMA-0
solution and the standard LMA-I solution are illustrated in
Fig.~\ref{fig:Pee} (bottom).  The curves represent
probabilities averaged over time and over the production region
inside the Sun for the $^8$B and $pep$ components of the solar
neutrino spectrum according to \cite{bp04}. The probabilities for
$^7$Be and $pp$ neutrinos, not shown, are very close (with less
than $\sim 7\%$ difference) to those for $^8$B and $pep$
respectively in the energy range of these neutrino fluxes. The
energy intervals relevant to the different spectral components are
also shown in the figure. The interval for $^8$B neutrinos is cut
from below at $E_\nu=6.5$ MeV; this approximately corresponds to
the threshold of $T\simeq 5$ MeV in the electron energy at the SNO
experiment.

Interestingly, the LMA-0 solution has the features sought
after in \cite{smirnovhomestake}, where a sterile neutrino was
introduced to eliminate the LMA-I upturn at SNO and improve the
agreement with the Homestake rate.

It is remarkable that, despite the wealth of data collected up to
this point, such radically different scenarios as LMA-0 and LMA-I
cannot be distinguished. The data expected in the next several
years, on the other hand, should be able to resolve the ambiguity.
First, if the SNO experiment lowers its energy threshold, it may
be able, with sufficient statistics, to look for the upturn
expected for the LMA-I solution. The absence of the upturn would
indicate the presence of NSI, or some other new physics.
Second, the expected $^7$Be flux in the case of LMA-0 is lower,
and the difference could be detected by the Borexino experiment
(or by the future solar phase of KamLAND). Third, the small value
of $\Delta m^2$ could be detected in the KamLAND spectrum data.
The predicted spectra for LMA-I (standard interactions) and LMA-0
are shown in Fig.~\ref{fig:Pee} (top). It can be seen that the two
are different at high energy where LMA-0 predicts
more events. Thus, to make the discrimination it is necessary to
both collect enough data and have a reliable calculation of the
antineutrino flux for $E_{\bar{\nu}}\gtrsim 6$ MeV. Finally, as evident from
Fig.~\ref{fig:Pee}, the two solutions make dramatically different
predictions for a \textsl{pep} experiment\footnote{It has been
proposed to detect pep neutrinos by electron scattering (M. Chen,
private communications) or by CC absorption (R. S. Raghavan, talk
at NOON04, Japan).}.

While an observation consistent with the standard LMA-I solution would
allow placing a very effective constraint on the neutrino-matter
interactions,
%excluding ranges of parameters that cannot be
%accessed by data from either present-day accelerator-based
%experiments or atmospheric neutrino experiments. At the same time,
a discovery of a deviation consistent with the NSI signal would
have truly profound particle physics implications. For example,
according to Refs.~\cite{BerezhianiRossi,Davidson:2003ha}, such
interaction could be due to the operator of the form
$M^{-4}\bar{l}_R(H^\dagger\vec{\sigma}L)(\bar{L}\vec{\sigma}H)l_R$
 $\propto v^2M^{-4}(\bar{\nu}\nu)(\bar{l}_R
l_R)$. For this operator to have an effect on the solar neutrino
survival probability, the coefficient $\propto v^2M^{-4}$ must not
be too small, {\it i.e.}, the scale of new physics $M$ must not be
much higher that the weak scale (Higgs vev $v$). Thus, by looking
for the NSI signatures in solar/reactor neutrinos the experiments
could in fact be probing new physics at the TeV scale.

In summary, the present-day loose bounds on some of the neutrino
interaction parameters introduce a serious uncertainty in the value of
$\Delta m^2$ extracted from solar and KamLAND data, allowing for a
new, disconnected solution. These uncertainties might be eliminated in
the next several years, as more data are collected and analyzed by
solar and KamLAND experiments. The constraints on the neutrino
interactions presented here will be further extended. On the other
hand, deviations from the SM neutrino interactions could indicate the
presence of radically new physics. We urge experimentalists to
consider these points in their data analysis.

A. F. was supported by the Department of Energy, under contract
W-7405-ENG-36; C. L. and C. P.-G. were supported by a grant-in-aid
from the W.M. Keck Foundation and the NSF PHY-0070928.


\begin{thebibliography}{33}
\expandafter\ifx\csname
natexlab\endcsname\relax\def\natexlab#1{#1}\fi
\expandafter\ifx\csname bibnamefont\endcsname\relax
  \def\bibnamefont#1{#1}\fi
\expandafter\ifx\csname bibfnamefont\endcsname\relax
  \def\bibfnamefont#1{#1}\fi
\expandafter\ifx\csname citenamefont\endcsname\relax
  \def\citenamefont#1{#1}\fi
\expandafter\ifx\csname url\endcsname\relax
  \def\url#1{\texttt{#1}}\fi
\expandafter\ifx\csname
urlprefix\endcsname\relax\def\urlprefix{URL }\fi
\providecommand{\bibinfo}[2]{#2}
\providecommand{\eprint}[2][]{\url{#2}}

\bibitem[{\citenamefont{Friedland et~al.}()\citenamefont{Friedland, Lunardini,
  and Pe\~na-Garay}}]{inprep}
\bibinfo{author}{\bibfnamefont{A.}~\bibnamefont{Friedland}},
  \bibinfo{author}{\bibfnamefont{C.}~\bibnamefont{Lunardini}},
  \bibnamefont{and}
  \bibinfo{author}{\bibfnamefont{C.}~\bibnamefont{Pe\~na-Garay}},
  \bibinfo{note}{in preparation}.

\bibitem{Roulet}
E.~Roulet,
 %``Mikheyev-Smirnov-Wolfenstein effect with flavor-changing neutrino
%nteractions,''
Phys.\ Rev.\ D {\bf 44}, 935 (1991).
%%CITATION = PHRVA,D44,935;%%

\bibitem[{\citenamefont{Barger et~al.}(1991)\citenamefont{Barger, Phillips, and
  Whisnant}}]{Barger:1991ae}
\bibinfo{author}{\bibfnamefont{V.~D.} \bibnamefont{Barger}},
  \bibinfo{author}{\bibfnamefont{R.~J.~N.} \bibnamefont{Phillips}},
  \bibnamefont{and} \bibinfo{author}{\bibfnamefont{K.}~\bibnamefont{Whisnant}},
  \bibinfo{journal}{Phys. Rev.} \textbf{\bibinfo{volume}{D44}},
  \bibinfo{pages}{1629} (\bibinfo{year}{1991}).

\bibitem[{\citenamefont{Fogli and Lisi}(1994)}]{Fogli:1994xv}
\bibinfo{author}{\bibfnamefont{G.~L.} \bibnamefont{Fogli}} \bibnamefont{and}
  \bibinfo{author}{\bibfnamefont{E.}~\bibnamefont{Lisi}},
  \bibinfo{journal}{Astropart. Phys.} \textbf{\bibinfo{volume}{2}},
  \bibinfo{pages}{91} (\bibinfo{year}{1994}).

\bibitem[{\citenamefont{Bergmann}(1998)}]{Bergmann:1998mr}
\bibinfo{author}{\bibfnamefont{S.}~\bibnamefont{Bergmann}},
  \bibinfo{journal}{Nucl. Phys.} \textbf{\bibinfo{volume}{B515}},
  \bibinfo{pages}{363} (\bibinfo{year}{1998}).

\bibitem{nutev}
G.~P.~Zeller {\it et al.}  [NuTeV Collaboration],
 %``A precise determination of electroweak parameters in neutrino nucleon
%scattering,''
Phys.\ Rev.\ Lett.\  {\bf 88}, 091802 (2002) [Erratum-ibid.\  {\bf
90}, 239902 (2003)] [arXiv:hep-ex/0110059].
%%CITATION = HEP-EX 0110059;%%

\bibitem{charm}
P.~Vilain {\it et al.}  [CHARM-II Collaboration],
 %``Precision measurement of electroweak parameters from the scattering of
%muon-neutrinos on electrons,''
Phys.\ Lett.\ B {\bf 335}, 246 (1994).
%%CITATION = PHLTA,B335,246;%%

\bibitem[{\citenamefont{Berezhiani and Rossi}(2002)}]{BerezhianiRossi}
\bibinfo{author}{\bibfnamefont{Z.}~\bibnamefont{Berezhiani}} \bibnamefont{and}
  \bibinfo{author}{\bibfnamefont{A.}~\bibnamefont{Rossi}},
  \bibinfo{journal}{Phys. Lett.} \textbf{\bibinfo{volume}{B535}},
  \bibinfo{pages}{207} (\bibinfo{year}{2002}), \eprint{hep-ph/0111137}.

\bibitem[{\citenamefont{Davidson et~al.}(2003)\citenamefont{Davidson,
  Pena-Garay, Rius, and Santamaria}}]{Davidson:2003ha}
\bibinfo{author}{\bibfnamefont{S.}~\bibnamefont{Davidson}},
  \bibinfo{author}{\bibfnamefont{C.}~\bibnamefont{Pe\~na-Garay}},
  \bibinfo{author}{\bibfnamefont{N.}~\bibnamefont{Rius}}, \bibnamefont{and}
  \bibinfo{author}{\bibfnamefont{A.}~\bibnamefont{Santamaria}},
  \bibinfo{journal}{JHEP} \textbf{\bibinfo{volume}{03}}, \bibinfo{pages}{011}
  (\bibinfo{year}{2003}), \eprint{hep-ph/0302093}.

\bibitem{W}
L.~Wolfenstein,
%``Neutrino Oscillations In Matter,''
Phys.\ Rev.\ D {\bf 17}, 2369 (1978).
%%CITATION = PHRVA,D17,2369;%%

\bibitem[{\citenamefont{Donnelly et~al.}(1974)\citenamefont{Donnelly, Hitlin,
  Schwartz, Walecka, and Wiesner}}]{Donnelly:1974pq}
\bibinfo{author}{\bibfnamefont{T.~W.} \bibnamefont{Donnelly}},
  \bibinfo{author}{\bibfnamefont{D.}~\bibnamefont{Hitlin}},
  \bibinfo{author}{\bibfnamefont{M.}~\bibnamefont{Schwartz}},
  \bibinfo{author}{\bibfnamefont{J.~D.} \bibnamefont{Walecka}},
  \bibnamefont{and} \bibinfo{author}{\bibfnamefont{S.~J.}
  \bibnamefont{Wiesner}}, \bibinfo{journal}{Phys. Lett.}
  \textbf{\bibinfo{volume}{B49}}, \bibinfo{pages}{8} (\bibinfo{year}{1974}).

\bibitem{3nu}
T.~K.~Kuo and J.~Pantaleone,
%``The Solar Neutrino Problem And Three Neutrino Oscillations,''
Phys.\ Rev.\ Lett.\  {\bf 57}, 1805 (1986);
%%CITATION = PRLTA,57,1805;%%
S.~Toshev,
%``Resonant Amplification Of Three Neutrino Oscillations In Matter,''
Phys.\ Lett.\ B {\bf 185}, 177 (1987) [Erratum-ibid.\ B {\bf 192},
478 (1987)].
%%CITATION = PHLTA,B185,177;%%

\bibitem[{\citenamefont{Apollonio et~al.}(1998)}]{Apollonio:1998xe}
\bibinfo{author}{\bibfnamefont{M.}~\bibnamefont{Apollonio}}
  \bibnamefont{et~al.},
  \bibinfo{journal}{Phys. Lett.} \textbf{\bibinfo{volume}{B420}},
  \bibinfo{pages}{397} (\bibinfo{year}{1998}).

\bibitem[{\citenamefont{Gonzalez-Garcia
  et~al.}(2001)\citenamefont{Gonzalez-Garcia, Grossman, Gusso, and
  Nir}}]{Gonzalez-Garcia:2001mp}
\bibinfo{author}{\bibfnamefont{M.~C.} \bibnamefont{Gonzalez-Garcia}},
  \bibinfo{author}{\bibfnamefont{Y.}~\bibnamefont{Grossman}},
  \bibinfo{author}{\bibfnamefont{A.}~\bibnamefont{Gusso}}, \bibnamefont{and}
  \bibinfo{author}{\bibfnamefont{Y.}~\bibnamefont{Nir}},
  \bibinfo{journal}{Phys. Rev.} \textbf{\bibinfo{volume}{D64}},
  \bibinfo{pages}{096006} (\bibinfo{year}{2001}), \eprint{hep-ph/0105159}.

\bibitem[{\citenamefont{Campanelli and Romanino}(2002)}]{Campanelli:2002cc}
\bibinfo{author}{\bibfnamefont{M.}~\bibnamefont{Campanelli}} \bibnamefont{and}
  \bibinfo{author}{\bibfnamefont{A.}~\bibnamefont{Romanino}},
  \bibinfo{journal}{Phys. Rev.} \textbf{\bibinfo{volume}{D66}},
  \bibinfo{pages}{113001} (\bibinfo{year}{2002}), \eprint{hep-ph/0207350}.

\bibitem[{\citenamefont{de~Gouvea et~al.}(2000)\citenamefont{de~Gouvea,
  Friedland, and Murayama}}]{darkside}
\bibinfo{author}{\bibfnamefont{A.}~\bibnamefont{de~Gouvea}},
  \bibinfo{author}{\bibfnamefont{A.}~\bibnamefont{Friedland}},
  \bibnamefont{and} \bibinfo{author}{\bibfnamefont{H.}~\bibnamefont{Murayama}},
  \bibinfo{journal}{Phys. Lett.} \textbf{\bibinfo{volume}{B490}},
  \bibinfo{pages}{125} (\bibinfo{year}{2000}), \eprint{hep-ph/0002064}.
  %%CITATION = HEP-PH 0002064;%%

\bibitem{Parke}
S.~J.~Parke,
%``Nonadiabatic Level Crossing In Resonant Neutrino Oscillations,''
Phys.\ Rev.\ Lett.\  {\bf 57}, 1275 (1986).
%%CITATION = PRLTA,57,1275;%%


\bibitem[{\citenamefont{Toshev}(1987)}]{Toshev1987}
\bibinfo{author}{\bibfnamefont{S.}~\bibnamefont{Toshev}},
  \bibinfo{journal}{Phys. Lett.} \textbf{\bibinfo{volume}{B196}},
  \bibinfo{pages}{170} (\bibinfo{year}{1987}).

\bibitem[{\citenamefont{Petcov}(1988)}]{Petcov1988}
\bibinfo{author}{\bibfnamefont{S.~T.} \bibnamefont{Petcov}},
  \bibinfo{journal}{Phys. Lett.} \textbf{\bibinfo{volume}{B200}},
  \bibinfo{pages}{373} (\bibinfo{year}{1988}).

\bibitem{MS}
S.~P.~Mikheev and A.~Y.~Smirnov,
 %``Resonance Enhancement Of Oscillations In Matter And Solar Neutrino
%Spectroscopy,''
Sov.\ J.\ Nucl.\ Phys.\  {\bf 42}, 913 (1985)
[Yad.\ Fiz.\  {\bf 42}, 1441 (1985)].
%%CITATION = SJNCA,42,913;%%



\bibitem[{\citenamefont{Friedland}(2001)}]{jumpinginsideout}
\bibinfo{author}{\bibfnamefont{A.}~\bibnamefont{Friedland}},
  \bibinfo{journal}{Phys. Rev.} \textbf{\bibinfo{volume}{D64}},
  \bibinfo{pages}{013008} (\bibinfo{year}{2001}).
  %%CITATION = HEP-PH 0010231;%%

\bibitem[{\citenamefont{Mansour and Kuo}(1998)}]{Mansour:1998fi}
\bibinfo{author}{\bibfnamefont{S.}~\bibnamefont{Mansour}} \bibnamefont{and}
  \bibinfo{author}{\bibfnamefont{T.~K.} \bibnamefont{Kuo}},
  \bibinfo{journal}{Phys. Rev.} \textbf{\bibinfo{volume}{D58}},
  \bibinfo{pages}{013012} (\bibinfo{year}{1998}), \eprint{hep-ph/9711424}.

\bibitem[{\citenamefont{Bahcall et~al.}(2002)\citenamefont{Bahcall,
  Gonzalez-Garcia, and Pena-Garay}}]{solarfit}
\bibinfo{author}{\bibfnamefont{J.~N.} \bibnamefont{Bahcall}},
  \bibinfo{author}{\bibfnamefont{M.~C.} \bibnamefont{Gonzalez-Garcia}},
  \bibnamefont{and}
  \bibinfo{author}{\bibfnamefont{C.}~\bibnamefont{Pe\~na-Garay}},
  \bibinfo{journal}{JHEP} \textbf{\bibinfo{volume}{07}}, \bibinfo{pages}{054}
  (\bibinfo{year}{2002}), \eprint{hep-ph/0204314}.

\bibitem[{\citenamefont{Bahcall et~al.}(2003)\citenamefont{Bahcall,
  Gonzalez-Garcia, and Pena-Garay}}]{klandfit}
\bibinfo{author}{\bibfnamefont{J.~N.} \bibnamefont{Bahcall}},
  \bibinfo{author}{\bibfnamefont{M.~C.} \bibnamefont{Gonzalez-Garcia}},
  \bibnamefont{and}
  \bibinfo{author}{\bibfnamefont{C.}~\bibnamefont{Pe\~na-Garay}},
  \bibinfo{journal}{JHEP} \textbf{\bibinfo{volume}{02}}, \bibinfo{pages}{009}
  (\bibinfo{year}{2003}), \eprint{hep-ph/0212147}.

\bibitem[{\citenamefont{Cleveland et~al.}(1998)}]{chlorine}
\bibinfo{author}{\bibfnamefont{B.~T.} \bibnamefont{Cleveland}}
  \bibnamefont{et~al.}, \bibinfo{journal}{Astrophys. J.}
  \textbf{\bibinfo{volume}{496}}, \bibinfo{pages}{505} (\bibinfo{year}{1998}).

\bibitem[{\citenamefont{Abdurashitov et~al.}(2002)}]{sage}
\bibinfo{author}{\bibfnamefont{J.~N.} \bibnamefont{Abdurashitov}}
  \bibnamefont{et~al.} (\bibinfo{collaboration}{SAGE}), \bibinfo{journal}{J.
  Exp. Theor. Phys.} \textbf{\bibinfo{volume}{95}}, \bibinfo{pages}{181}
  (\bibinfo{year}{2002}), \eprint{astro-ph/0204245}.

\bibitem[{\citenamefont{Hampel et~al.}(1999)}]{Hampel:1998xg}
\bibinfo{author}{\bibfnamefont{W.}~\bibnamefont{Hampel}} \bibnamefont{et~al.},
  \bibinfo{journal}{Phys. Lett.}
  \textbf{\bibinfo{volume}{B447}}, \bibinfo{pages}{127} (\bibinfo{year}{1999}).

\bibitem[{\citenamefont{Kirsten}()}]{gno}
\bibinfo{author}{\bibfnamefont{T.}~\bibnamefont{Kirsten}}, \bibinfo{note}{talk
  at the XXth International Conference on Neutrino Physics and Astrophysics
  (NU2002)}.

\bibitem[{\citenamefont{Fukuda et~al.}(2002)}]{sksollast}
\bibinfo{author}{\bibfnamefont{S.}~\bibnamefont{Fukuda}} \bibnamefont{et~al.}
  (\bibinfo{collaboration}{Super-Kamiokande}), \bibinfo{journal}{Phys. Lett.}
  \textbf{\bibinfo{volume}{B539}}, \bibinfo{pages}{179} (\bibinfo{year}{2002}),
  \eprint{hep-ex/0205075}.

\bibitem[{\citenamefont{Ahmad et~al.}(2001)}]{Ahmad:2001an}
\bibinfo{author}{\bibfnamefont{Q.~R.} \bibnamefont{Ahmad}} \bibnamefont{et~al.}
  (\bibinfo{collaboration}{SNO}), \bibinfo{journal}{Phys. Rev. Lett.}
  \textbf{\bibinfo{volume}{87}}, \bibinfo{pages}{071301}
  (\bibinfo{year}{2001}), \eprint{nucl-ex/0106015}.

\bibitem[{\citenamefont{Ahmad et~al.}(2002{\natexlab{a}})}]{Ahmad:2002jz}
\bibinfo{author}{\bibfnamefont{Q.~R.} \bibnamefont{Ahmad}} \bibnamefont{et~al.}
  (\bibinfo{collaboration}{SNO}), \bibinfo{journal}{Phys. Rev. Lett.}
  \textbf{\bibinfo{volume}{89}}, \bibinfo{pages}{011301}
  (\bibinfo{year}{2002}{\natexlab{a}}), \eprint{nucl-ex/0204008}.

\bibitem[{\citenamefont{Ahmad et~al.}(2002{\natexlab{b}})}]{Ahmad:2002ka}
\bibinfo{author}{\bibfnamefont{Q.~R.} \bibnamefont{Ahmad}} \bibnamefont{et~al.}
  (\bibinfo{collaboration}{SNO}), \bibinfo{journal}{Phys. Rev. Lett.}
  \textbf{\bibinfo{volume}{89}}, \bibinfo{pages}{011302}
  (\bibinfo{year}{2002}{\natexlab{b}}), \eprint{nucl-ex/0204009}.

\bibitem{snoII}
S.~N.~Ahmed {\it et al.}  (SNO),
%``Measurement of the total active B-8 solar neutrino flux at the Sudbury
%Neutrino Observatory with enhanced neutral current sensitivity,''
Phys.\ Rev.\ Lett.\  {\bf 92}, 181301 (2004), nucl-ex/0309004.
%%CITATION = NUCL-EX 0309004;%%

\bibitem{Vogel2002}
The experimental constraints on this cross
  section are summarized in 
M.~Butler, J.~W.~Chen and P.~Vogel,
%``Constraints on two-body axial currents from reactor antineutrino  deuteron
%breakup reactions,''
Phys.\ Lett.\ B {\bf 549}, 26 (2002)
[arXiv:nucl-th/0206026].
%%CITATION = NUCL-TH 0206026;%%

\bibitem[{\citenamefont{Bahcall and Pinsonneault}(2004)}]{bp04}
J.~N.~Bahcall and M.~H.~Pinsonneault,
%``What do we (not) know theoretically about solar neutrino fluxes?,''
Phys.\ Rev.\ Lett.\  {\bf 92}, 121301 (2004);
%%CITATION = ASTRO-PH 0402114;%%
  http://www.sns.ias.edu/$\sim$jnb.

\bibitem[{\citenamefont{Eguchi et~al.}(2003)}]{kamland}
\bibinfo{author}{\bibfnamefont{K.}~\bibnamefont{Eguchi}} \bibnamefont{et~al.}
  (\bibinfo{collaboration}{KamLAND}), \bibinfo{journal}{Phys. Rev. Lett.}
  \textbf{\bibinfo{volume}{90}}, \bibinfo{pages}{021802}
  (\bibinfo{year}{2003}), \eprint{hep-ex/0212021}.

\bibitem{verylowlma}
C.~Burgess, N.~S.~Dzhalilov, M.~Maltoni, T.~I.~Rashba,
V.~B.~Semikoz, M.~Tortola and J.~W.~F.~Valle,
%``Large mixing angle oscillations as a probe of the deep solar interior,''
Astrophys.\ J.\  {\bf 588}, L65 (2003) [arXiv:hep-ph/0209094];
%%CITATION = HEP-PH 0209094;%%
M.~M.~Guzzo, P.~C.~de Holanda and N.~Reggiani,
 %``Large mixing angle solution to the solar neutrino problem and random  matter
%density perturbations,''
Phys.\ Lett.\ B {\bf 569}, 45 (2003) [arXiv:hep-ph/0303203].
%%CITATION = HEP-PH 0303203;%%


\bibitem{Loreti1994}
F.~N.~Loreti and A.~B.~Balantekin,
%``Neutrino oscillations in noisy media,''
Phys.\ Rev.\ D {\bf 50}, 4762 (1994) [arXiv:nucl-th/9406003].
%%CITATION = NUCL-TH 9406003;%%

\bibitem[{\citenamefont{Fornengo et~al.}(2002)\citenamefont{Fornengo, Maltoni,
  Bayo, and Valle}}]{Fornengo:2001pm}
\bibinfo{author}{\bibfnamefont{N.}~\bibnamefont{Fornengo}},
  \bibinfo{author}{\bibfnamefont{M.}~\bibnamefont{Maltoni}},
  \bibinfo{author}{\bibfnamefont{R.~T.} \bibnamefont{Bayo}}, \bibnamefont{and}
  \bibinfo{author}{\bibfnamefont{J.~W.~F.} \bibnamefont{Valle}},
  \bibinfo{journal}{Phys. Rev.} \textbf{\bibinfo{volume}{D65}},
  \bibinfo{pages}{013010} (\bibinfo{year}{2002}), \eprint{hep-ph/0108043}.

\bibitem[{\citenamefont{Guzzo et~al.}(2002)}]{brazilians}
\bibinfo{author}{\bibfnamefont{M.}~\bibnamefont{Guzzo}} \bibnamefont{et~al.},
  \bibinfo{journal}{Nucl. Phys.} \textbf{\bibinfo{volume}{B629}},
  \bibinfo{pages}{479} (\bibinfo{year}{2002}).

\bibitem[{\citenamefont{de~Holanda and Smirnov}(2003)}]{smirnovhomestake}
\bibinfo{author}{\bibfnamefont{P.~C.} \bibnamefont{de~Holanda}}
  \bibnamefont{and} \bibinfo{author}{\bibfnamefont{A.~Y.}
  \bibnamefont{Smirnov}} (\bibinfo{year}{2003}), \eprint{hep-ph/0307266}.

\bibitem[{\citenamefont{Friedland and Lunardini}(2003{\natexlab{a}})}]{nunu1}
\bibinfo{author}{\bibfnamefont{A.}~\bibnamefont{Friedland}} \bibnamefont{and}
  \bibinfo{author}{\bibfnamefont{C.}~\bibnamefont{Lunardini}},
  \bibinfo{journal}{Phys. Rev.} \textbf{\bibinfo{volume}{D68}},
  \bibinfo{pages}{013007} (\bibinfo{year}{2003}{\natexlab{a}});
 \bibinfo{journal}{JHEP} \textbf{\bibinfo{volume}{10}}, \bibinfo{pages}{043}
  (\bibinfo{year}{2003}{\natexlab{b}}).
  %%CITATION = HEP-PH 0307140;%%
  %%CITATION = HEP-PH 0304055;%%

\bibitem{Burgess2003}
C.~P.~Burgess, N.~S.~Dzhalilov, M.~Maltoni, T.~I.~Rashba,
V.~B.~Semikoz, M.~A.~Tortola and J.~W.~F.~Valle,
%``Cornering solar radiative-zone fluctuations with KamLAND and SNO salt,''
JCAP {\bf 0401}, 007 (2004) [arXiv:hep-ph/0310366].
%%CITATION = HEP-PH 0310366;%%


\end{thebibliography}
\end{document}